\documentstyle[11pt,aaspp4]{article}

\begin{document}
\tighten
\title{The Physical Characteristics of the Small-Scale Interstellar Structure
	towards $\mu$ Crucis\footnote{Based on observations with the
	NASA/ESA {\it Hubble Space Telescope}, obtained at the Space
	Telescope Science Institute, which is operated by the Association
	for Research in Astronomy, Inc., under NASA contract NAS5-26555.}}
\author{J. T. Lauroesch, David M. Meyer, and John K. Watson}
\affil{Department of Physics and Astronomy}
\affil{Northwestern University}
\authoraddr{Evanston, Illinois 60208}
\authoremail{jtl@elvis.astro.nwu.edu, meyer@elvis.astro.nwu.edu,
	jkwatson@elvis.astro.nwu.edu}
\and
\author{J.C. Blades}
\affil{Space Telescope Science Institute}
\authoraddr{3700 San Martin Drive}
\authoraddr{Baltimore, MD 21218}
\authoremail{blades@stsci.edu}

\medskip\medskip
\centerline{\large To appear in the {\it Astrophysical Journal (Letters)}}
\received{July 23, 1998}
\accepted{September 3, 1998}
\sluginfo

\begin{abstract}
We present {\it HST}/GHRS echelle observations of multiple interstellar lines
of \ion{C}{1}, \ion{Mg}{1}, \ion{Cr}{2}, and \ion{Zn}{2} towards both stars
in the $\mu$ Cru binary system.  Despite large differences in the profiles
of the neutral species, no significant variations between the stars
are seen in the \ion{Cr}{2} and \ion{Zn}{2} line profiles.  In particular, the
\ion{Zn}{2} absorption observed at -8.6 km/sec towards $\mu$ Cru is constant
despite greatly enhanced columns of the neutral species at this velocity
towards $\mu^1$ Cru.  An analysis of the fine-structure excitation of
\ion{C}{1} in this cloud implies that the density is $\rm n_H < 250\ cm^{-3}$.
From the lack of variation in the (optical)
\ion{Na}{1} D2 line profiles towards $\mu^1$ and $\mu^2$ Cru in spectra taken
21 months apart, we can place a lower limit to the size of the structures
of $\sim$10 AU.  These results are discussed in the context of recent
radio and optical studies of apparently pervasive high density small-scale
interstellar structure.
\end{abstract}


\clearpage

\section{Introduction}

There is growing evidence from recent radio and optical observations
that the diffuse interstellar medium (ISM) exhibits significant
subparsec-scale variations down to limits of a few AU (Frail {\it et al}.\ 
1994; Davis {\it et al}.\ 1996; Meyer \& Blades 1996; Watson \& Meyer 1996).
Utilizing multi-epoch observations of 21 cm absorption towards high-velocity
pulsars, Frail {\it et al}.\ find pervasive variations in the \ion{H}{1}
opacity on scales of 5 to 100 AU with implied densities of
$\rm n_H\sim 10^4-10^5\ cm^{-3}$.  Probing larger scales (500-30,000 AU)
through optical observations of the interstellar \ion{Na}{1} absorption
towards a number of resolved binary star systems, Watson \& Meyer (1996)
also found a ubiquitous pattern of small interstellar structures and
inferred densities of $\rm n_H\ga 10^3\ cm^{-3}$ in some structures.  Due
to their large overpressures with respect to the intercloud medium,
such structures cannot be accommodated in any abundance in the standard
McKee \& Ostriker (1977) model of the ISM.  Possible explanations for
these structures range from small-scale filamentary geometries (Heiles 1997),
to fractal geometries driven by turbulence (Elmegreen 1997), to a separate
population of small, dense self-gravitating clouds (Walker \& Wardle 1998).

An important point discussed by Heiles (1997) is that the high volume
densities of the small-scale structures have not been measured directly --
they have been inferred by dividing the observed column density differences
by the sampled transverse length scales.  Furthermore, in the case of the
\ion{Na}{1} observations, since \ion{Na}{1} is not the dominant ion
in \ion{H}{1} clouds it may not be a reliable tracer of \ion{H}{1}
column density differences, especially at low column densities (Welty,
Hobbs, \& Kulkarni 1994).  In this {\it Letter}, we present {\it HST} GHRS
echelle observations of the interstellar absorption for \ion{C}{1},
\ion{Mg}{1}, \ion{Cr}{2}, and \ion{Zn}{2} towards both members of the
binary $\mu$ Cru (Meyer \& Blades 1996) in the
first effort to apply UV interstellar diagnostics towards an understanding
of the physical conditions in the structures giving rise to the observed
small-scale variations.  In addition, we present new high-resolution
\ion{Na}{1} observations of this binary that afford a temporal probe of
the ISM structure at a scale similar to that sampled by
Frail {\it et al}.\ (1994).

\section{Observations}

\subsection{{\it HST}/GHRS Observations}
 
The $\mu$ Cru binary system consists of a B2IV-V and a B5Ve star at a
distance of 170 pc with a separation of 38{\arcsec}.8 (corresponding to
a projected linear separation of 6600 AU or 0.03 pc).  Both stars were
observed using the {\it HST} Goddard High Resolution Spectrograph
(GHRS).  Four wavelength intervals were covered using the high resolution
echelle gratings (ECH-A near 1271\AA, and ECH-B near 2026, 2062,
and 2852\AA) to observe lines of \ion{C}{1}, \ion{Mg}{1}, \ion{Cr}{2},
and \ion {Zn}{2} at a resolution of $\sim$3.5 km/sec.  The data were
taken in FP-SPLIT mode, the calibrated data files as delivered by STScI
were coadded in the usual fashion using the STSDAS tasks {\tt poffsets,
dopoff,} and {\tt specalign}.  For the rapidly rotating star $\mu^2$ Cru the
continuum was fitted using low-order polynomials.  For $\mu^1$ Cru continuum
fitting was more difficult due to the low $v\cdot sin(i)$ (48 km/sec).
Here smaller regions of the spectrum were fitted using
higher-order polynomials; the resulting profile for each line was then
compared to the other line of the same species towards $\mu^1$ Cru (and
the same line towards $\mu^2$ Cru) to be sure that the continuum was
properly placed and that the entire interstellar line was excluded from the
fit.  The resulting profiles are shown in Figure~1; note that
the variation is largely confined to the lines of the neutral species,
with no significant variations in the \ion{Cr}{2} or \ion{Zn}{2} profiles.

\subsection{Anglo-Australian Telescope/UHRF Observations}

The interstellar \ion{Na}{1} D2 absorption was re-observed with the
Ultra-High Resolution Facility (UHRF) on the Anglo-Australian Telescope (AAT)
towards both members of the $\mu$ Cru binary star system in March 1997,
21 months after the first observations reported by Meyer \& Blades (1996).
A Tek 1024$\times$1024 CCD was used
for these observations; the data was then processed using a combination of
Figaro\footnote{Figaro is distributed as part of the Starlink Project,
which is run by CCLRC on behalf of PPARC.}
and IRAF\footnote{IRAF is distributed by the National Optical Astronomy
Observatories, which are operated by the Association of Universities for
Research in Astronomy, Inc., under cooperative agreement with the National
Science Foundation.} routines.  The spectrograph configuration for these
observations is the same as that described in Meyer \& Blades (1996); the
resulting resolution was $\sim 10^6$.

According to the {\it Hipparcos} catalog, the proper motion of these stars
is $\sim 0.034$\arcsec /year (ESA 1997), corresponding to a projected change in
position of $\sim 10$ AU over the 21 month period between observations.  As
shown in Figure~2, the line profiles are virtually identical; in addition, the
column densities, b-values, and relative velocities of the various
components derived by profile fitting\footnote{The
{\it fits6p} program was used to fit the data, this program is a somewhat
modified descendant of the program described by Welty, Hobbs, \& York (1991)}
are well within the calculated errors.  The striking lack of change in
the \ion{Na}{1} D2 profile implies a minimum cloudlet size of a few tens
of AU.  This result is not necessarily inconsistent with the
$\sim 5-100$ AU structures found by Frail {\it et al}.\ (1994), but
suggests that further monitoring of this sight-line would be worthwhile.

\section{Discussion}

The narrow, hyperfine-split absorption observed in \ion{Na}{1} at
-8.6 km/sec towards $\mu^1$ Cru provides an opportunity to derive the
physical conditions in a relatively isolated component.  In particular, the
components at 4.4 and 5.0 km/sec (which appear to be strongly enhanced
towards $\mu^2$ Cru) are severely blended at the resolution of the GHRS echelle
gratings, and thus any interpretation of these components is more ambiguous.
Profile fitting was used to derive the column densities listed in Table~1 for
the -8.6 km/sec component.  The most remarkable aspect of these numbers is
the enhancement over typical ISM values of the columns of the neutral
species relative to the observed \ion{Zn}{2} towards $\mu^1$ Cru,
despite the similarity of the \ion{Zn}{2} columns.
Since \ion{Zn}{2} is the dominant ion in typical \ion{H}{1} clouds, and
Zn is relatively undepleted (Sembach {\it et al}. 1995; Roth \& Blades 1995),
it is a direct measure of the \ion{H}{1} column -- the lack of variation
in the \ion{Zn}{2} profile implies that the variation in \ion{Na}{1} and
the other neutrals is occurring without a change in the \ion{H}{1} column.
These observations suggest that we may be seeing a small ``core'' in
the cloud at -8.6 km/sec where the neutrals vary, but where any variation
in the \ion{Zn}{2} (and hence \ion{H}{1}) column is masked by a broader
component with no associated neutral gas.  We will now explore the
physical conditions in this cloud, as well as several potential explanations
for the variations, and show that small-scale density and
temperature fluctuations in a diffuse interstellar cloud are the most
likely explanation for the observed neutral enhancements towards $\mu$ Cru.

Since $\mu^2$ Cru is a known Be star, one must consider the
possibility that the variations in absorption line profiles arise in
circumstellar gas around one or both members of the system, but there
is strong evidence that this gas is indeed interstellar in nature.
The narrow, hyperfine split \ion{Na}{1} components observed to vary between
the members of the $\mu$ Cru system arise in cold gas, and it is difficult
to imagine an environment near these B stars in which such cold clumps would
arise.  The lack of time-variability in the D2 absorption profiles also
argues against a circumstellar explanation since
an expanding circumstellar shell gas would be likely to shift or vary in
strength over this period.  Finally, the lack of strong \ion{C}{1}$^\star$
(and \ion{C}{1}$^{\star\star}$) absorption in the -8.6 km/sec component
implies that the radiation field cannot be much larger than $\sim 20$ times
the diffuse ultraviolet background (Jenkins \& Shaya 1979); calculations
show that the observed N(\ion{C}{1}$^\star$)/N(\ion{C}{1}$_{total}$) ratio
of $<$ 0.25 requires that this structure lies at least 1 pc from the B2V
star $\mu^1$ Cru.

We can estimate the hydrogen column density, and hence the density for this
component using the measured uncertainties in the Zn column (towards both
members of the system) as a rough upper limit to the column of \ion{Zn}{2}
in the structure giving rise to the \ion{Na}{1} variation.  We place a
2$\sigma$ upper limit on any enhancement at -8.6 km/sec of \ion{Zn}{2}
towards $\mu^1$ Cru of $\rm 4\times 10^{10}\ cm^{-3}$.  For typical
interstellar depletions this corresponds to
N(\ion{H}{1})$\rm < 2.5\times 10^{18}\ cm^{-3}$
(Sembach {\it et al}. 1995; Roth \& Blades 1995), which is less than
40\% of the value one would estimate from the observed \ion{Na}{1} column.
Given the binary separation of 6600 AU, this implies that the observed
variation may arise in a structure with a density contrast of
$\rm n_H < 25\ cm^{-3}$, which is appreciably less than that estimated in
previous studies of small-scale ISM structure.  However, note that 6600 AU is
only an upper limit to the size of the structure, and a smaller structure would
have a higher density.  Therefore, we now refine this simple argument
using the available data for the neutral species to directly estimate the
density without assuming a characteristic length scale.

First, we can place an upper limit on the hydrogen density using the
fine-structure equilibrium of \ion{C}{1}.  Given the measured ratio of
N(\ion{C}{1}$^\star$)/N(\ion{C}{1}$_{total}$)$<$0.25, we can place a strict
upper limit to the density for cloud temperatures $> 20$~K of
$\rm n_{H} < 250\ cm^{-3}$ (Jenkins \& Shaya 1979).  For a more typical
interstellar cloud temperature of $\sim 100$~K the density would be
$\rm n_{H} < 50\ cm^{-3}$, which is (again) substantially lower than the
densities estimated for small-scale structures in previous studies.  This
limit is independent of any estimate of the cloud size.

The measured columns of the neutral species can be combined with ionization
and recombination coefficients to derive estimates of the density in
this structure.  If we take \ion{Zn}{2} as a surrogate for the (unobserved)
first ions of carbon, sodium, and magnesium, we can estimate the
required columns by assuming Solar abundance ratios and a variety
of depletions (ranging from no depletion to typical cold cloud values).
P\'{e}quignot \& Aldrovandi (1986) have calculated ionization
and recombination coefficients as a function of temperature and electron
density for four choices of radiation field, including that of
Witt \& Johnson (1973).  For a temperature of 100~K and an assumed
$n_e/n_H$ ratio of $2\times 10^{-4}$ (from the ionization of carbon), we
can estimate $n_H$ using our observed columns (see Table~1).
First, to minimize the neutral fraction for C, Na, and Mg we will use
the total observed \ion{Zn}{2} column at -8.6 km/sec in these estimates.
Then, for no depletion, we derive $n_H < 205$, 120, and 190 $\rm cm^{-3}$ from
\ion{C}{1}, \ion{Na}{1}, and \ion{Mg}{1} respectively.  For a typical cold
cloud depletion pattern (Savage \& Sembach 1996), we estimate densities
of $\rm n_H < 260$, 240, and 620 $\rm cm^{-3}$ respectively.  Note that the
densities estimated from \ion{C}{1} and \ion{Na}{1} are consistent in both
cases, but that the density is apparently over-estimated for \ion{Mg}{1}
if we assume cold cloud depletions.  If the upper limit of
N(\ion{Zn}{2})$\rm < 4\times 10^{10}\ cm^{-3}$ estimated above is used,
densities of $\rm n_H < 730$, 425, and 540 $\rm cm^{-3}$ are found 
assuming no dust depletion.  These density estimates imply cloud sizes
of $\sim$ 350 -- 4,000 AU, which are comparable to the cloud size limits
of 10 and 6600 AU set by the optical observations.  The agreement
between the estimated cloud sizes and the observed limits is suggestive,
but note that there is an implicit assumption that there is little or no
ionized hydrogen in this structure.  If some fraction of the hydrogen is
ionized the density in the structure will be lower; however, as the ionized
fraction increases, the hydrogen column in the structure must decrease so
as not to exceed the limit on the cloud size set by the binary separation.

Now in addition to being density sensitive, the ratios of the columns of
neutrals to first ions is also sensitive to the temperature.  For
\ion{C}{1}, \ion{Na}{1}, and \ion{Mg}{1} the column density goes
roughly as $\rm\sim T^{-0.65}$ (P\'{e}quignot \& Aldrovandi 1986), thus a
difference in temperature of a factor of 100 would lead to a difference
in the columns of the neutrals of a factor of $\sim 20$, a value consistent
with the observed upper limit for \ion{Na}{1} at this velocity towards
$\mu^2$ Cru.  However, the profile fits to the \ion{Zn}{2} lines towards
both stars yield similar b-values for both lines-of-sight suggesting that the
temperature of the bulk of the material in these clouds is similar and
of order a few hundred degrees K (given $\rm b\sim 0.4$--$0.5$ km/sec),
although the presence of a somewhat narrower colder ``core'' towards
$\mu^1$ Cru cannot be ruled out.  Since the 21 cm opacity scales as
$\rm T^{-1}$, it seems at first glance that temperature fluctuations alone
could explain the observed variation in both the neutral and the 21 cm lines.
But as noted by Heiles (1997), such extreme temperature fluctuations
are difficult to understand for a number of reasons.  This suggests
that temperature fluctuations are likely to play at least some role
in producing the observed neutral variations, although it is unlikely to
be the sole mechanism involved.  Thus, a plausible model for the
absorbing clumps is that we are detecting a small, somewhat denser and
colder core towards $\mu^1$ Cru that is embedded in a larger cloud seen
towards both stars.

There are a number of similarities and differences between the properties of
the -8.6 km/sec cloudlet towards $\mu^1$ Cru and the structures observed in
radio studies.  The most direct comparisons can be made with the pulsar
study of Frail {\it et al}. (1994) as it is also a multi-epoch study
of absorption against a background point-source.  The scales sampled
in these studies are similar, thus the lack of variation observed towards
both members of the $\mu$ Cru system is somewhat surprising given the
ubiquity of the variations observed in the pulsar study, but is not
inconsistent with the inferred cloud sizes.  The lack of variation in
the observed \ion{Zn}{2} profiles between the stars suggests that
there is little or no variation in the total hydrogen column density towards
$\mu$ Cru\footnote{Note that the upper limit to the density from the
fine-structure equilibrium of \ion{C}{1} and the relatively low columns of
the observed neutral species imply that \ion{Zn}{1} cannot be the dominant
ionization stage of Zn in this velocity interval.},
which appears to be at odds with the large \ion{H}{1} variations observed in
the radio studies.  The upper limit for the density in this structure is
significantly different from the values estimated from the 21 cm observations,
which typically imply much larger densities in (perhaps) smaller clouds.
One explanation is that perhaps this gas is associated with one of
the Heiles (1997) filaments or sheets observed roughly ``face-on'', our
estimate for the size of the structure would then be an estimate
for the thickness of the filament/sheet (with the width of the structure being
constrained by the 6600 AU binary separation).  A large variation in the
\ion{Zn}{2} column density could be measured if the
filament/sheet was along the line-of-sight instead of perpendicular to it.
This sort of density fluctuation is also consistent with a fractal
ISM model (such as those of Elmegreen 1997), since density as well as
column density inhomogeneities can be a natural outcome of turbulent processes.

What, if any, fraction of the cloudlets detected as \ion{Na}{1} variations
are associated with the structures responsible for the 21 cm variations
towards radio sources is still uncertain.  Further observations of multiple
star systems with variable \ion{Na}{1} (in both lightly and more highly
reddened sight-lines) using {\it HST}/STIS and {\it FUSE} are
required for a fuller understanding of the physical conditions in the
\ion{Na}{1} variable components, and to (potentially) identify
components which show large \ion{H}{1} column density differences similar
to the 21 cm studies.  If no such components are found, then this suggests
that the \ion{Na}{1} and 21 cm studies are identifying different structures.

\acknowledgments
It is a pleasure to acknowledge several valuable conversations with the
late Lyman Spitzer, with Mark Wardle about extreme scattering events,
and with Dan Welty about the ionization equilibrium of neutral species.
In addition, we would like to thank the anonymous referee for 
their suggestions.

\clearpage

\begin{figure}
\caption{Optical (AAT) and ultraviolet (GHRS) lines observed towards both
        members of the binary system $\mu$ Cru, where the solid
        profile is that observed towards $\mu^1$ Cru, and the dashed is
        that observed towards $\mu^2$ Cru.  Note the relatively
        large differences between the lines-of-sight in the neutral
        species, compared to the slight differences seen for the first ions.}
\end{figure}
 
\begin{figure}
\caption{Comparison of the \ion{Na}{1} D$_2$ line profiles observed
        21 months apart towards both components of the binary
        system $\mu$ Cru.  Given the proper motion of $\mu$ Cru, the lack
        of variation in these profiles implies that the responsible
        structures are appreciably larger than 10 AU in size.}
\end{figure}

\begin{table}
\begin{center}
\caption{Column Densities for the -8.6 km/sec Component towards $\mu^1$ Cru}
\tablewidth{0pt}
\begin{tabular}{lccc}
\tableline\tableline
 Ion & Vacuum	 		& Oscillator			& Column     \\
     & Wavelength		& Strength\tablenotemark{a}	& Density    \\
     & (\AA)\tablenotemark{a}	&			&
	($\rm cm^{-2}$)\tablenotemark{b} \\ \tableline
\ion{C}{1}  & 1277.2454 & 0.09665 & 8.0$\pm 1.1\times 10^{11}$ \\
\ion{C}{1}$^{\star}$ & 1277.2823 & 0.07249 & $<2.7\times 10^{11}$ \\
\ion{C}{1}$^{\star\star}$ & 1277.5496 & 0.08117 & $<2.7\times 10^{11}$ \\
\ion{Na}{1} & 5891.5833 & 0.6550 & 6.78$\pm 0.08\times 10^{10}$ \\
            & 5897.5581 & 0.3270 & 6.94$\pm 0.15\times 10^{10}$ \\
\ion{Mg}{1} & 2026.4768 & 0.1154 & 4.1$\pm 0.5\times 10^{11}$   \\
            & 2852.9642 & 1.830 & 4.2$\pm 1.1\times 10^{11}$   \\
\ion{Cr}{2} & 2062.234 & 0.777\tablenotemark{c} & $<1.4\times 10^{11}$          \\
\ion{Zn}{2} & 2026.136 & 0.5150 & 1.4$\pm 0.3\times 10^{11}$    \\
            & 2062.664 & 0.2529 & 1.3$\pm 0.4\times 10^{11}$    \\ \hline
\end{tabular}
\tablenotetext{a}{From Morton (1991) unless noted otherwise.}
\tablenotetext{b}{Column densities derived from profile fitting the
	interstellar absorption corresponding to the -8.6 km/sec component
	towards $\mu^1$ Cru.  The errors quoted are 1$\sigma$, with
	2$\sigma$ upper limits given for the lines of
	\ion{C}{1}$^{\star}$, \ion{C}{1}$^{\star\star}$, and \ion{Cr}{2}.}
\tablenotetext{c}{Oscillator strength from Bergeson \& Lawler (1993).}
\end{center}
\end{table}

\clearpage

\begin{figure}[h]
\plotone{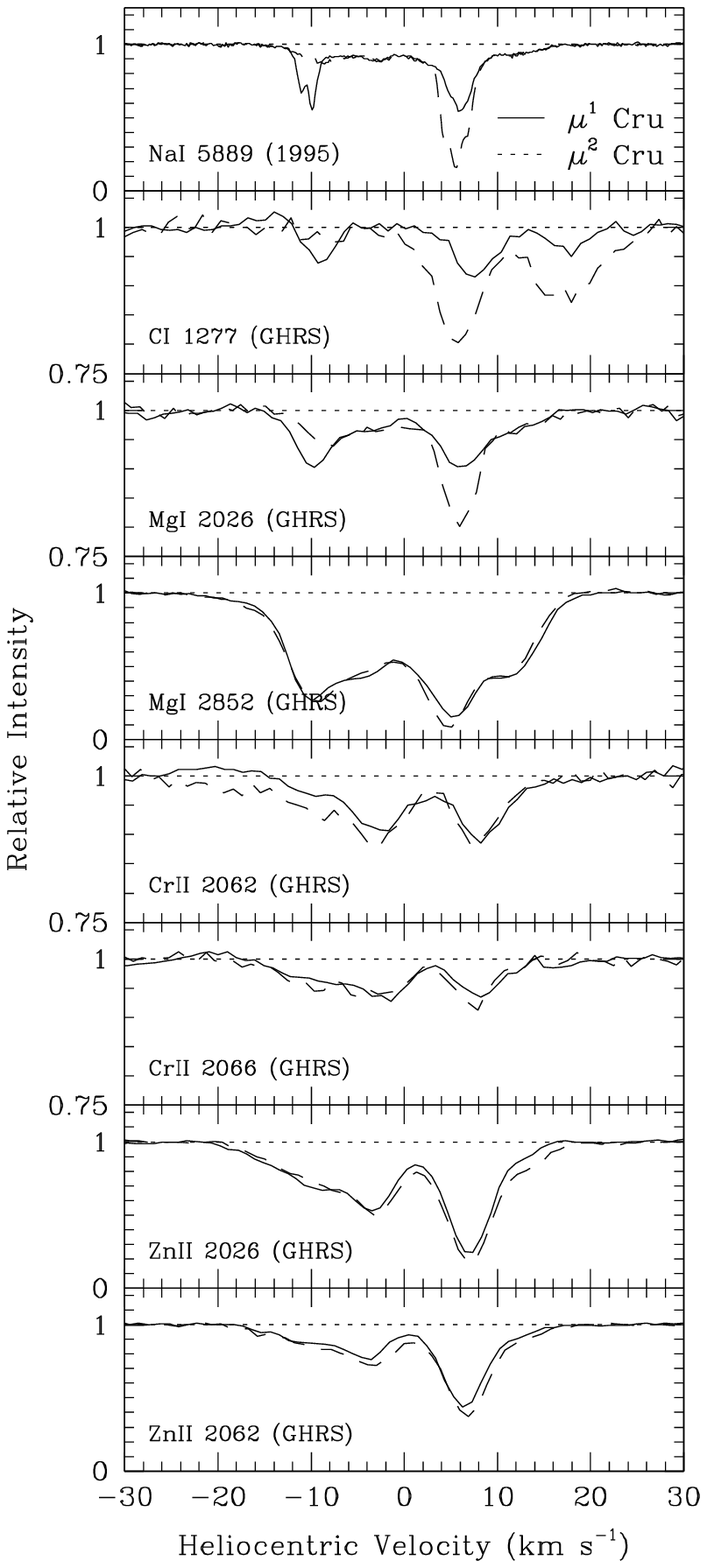}
\end{figure}
\centerline{\bf Figure 1}

\clearpage

\begin{figure}[h]
\plotone{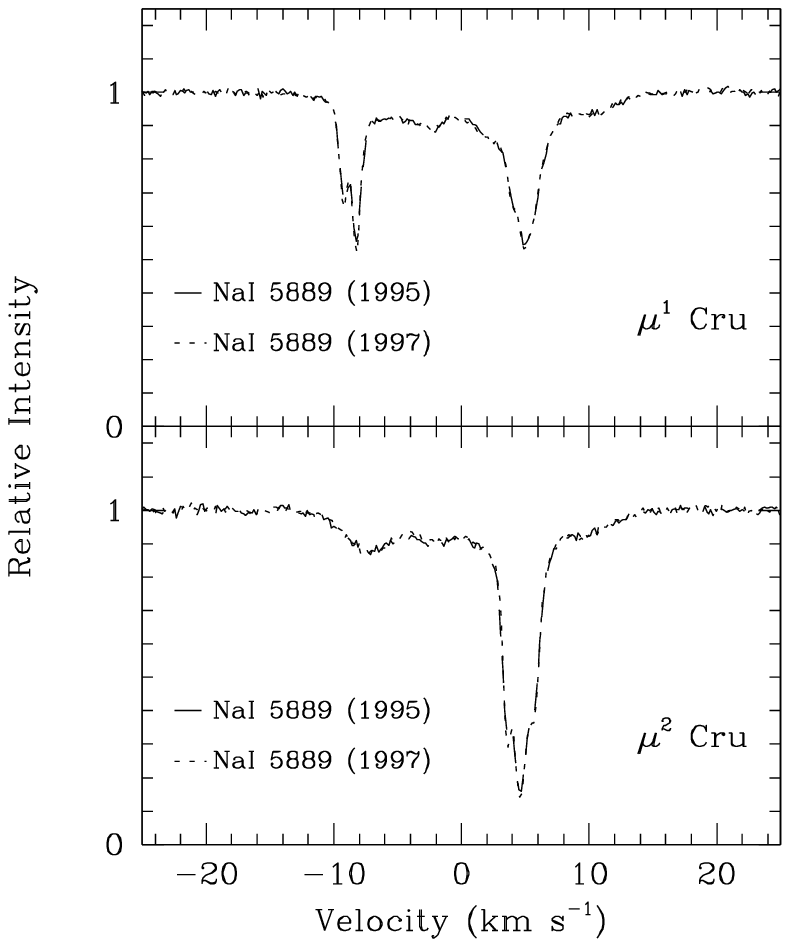}
\end{figure}
\centerline{\bf Figure 2}

\end{document}